\documentclass{sig-alternate-10pt}
\usepackage[margin=1in]{geometry} 
\usepackage[utf8]{inputenc}
\title{Systems for Near Real-Time Analysis of Large-Scale Dynamic Graphs}
\author{Luis M. Vaquero, Felix Cuadrado, Matei Ripeanu}

\begin{document}

\maketitle

\begin{abstract}

Graphs are widespread data structures used to model a wide variety of problems. The sheer amount of data to be processed has prompted the creation of a myriad of systems that help us cope with massive scale graphs. The pressure to deliver fast responses to queries on the graph is higher than ever before, as it is demanded by many applications (e.g. online recommendations, auctions, terrorism protection, etc.). In addition, graphs change continuously (so do the real world entities that typically represent). Systems must be ready for both: near real-time and dynamic masive graphs. We survey systems taking their scalability, real-time potential and capability to support dynamic changes to the graph as driving guidelines. The main techniques and limitations are distilled and categorised. The algorithms run on top of graph systems are not ready for prime time dynamism either. Therefore, a short overview on dynamic graph algorithms has also been included. 
\end{abstract}

\section{Introduction}

Graphs are simple data structures that help us model a wide variety of problems. They are broadly used, being key elements in our understanding of science (genetics, proteomics, computer networks or social sciences); and of industries, like network operators, travel, online dating, marketing companies or hospitals. 

These organisations store more data in the hope to better understand their own processes and achieve their goals. Graphs promise ease to model complicated scenarios and applicability in a variety of application domains. Consequently, a wide variety of large-scale graph processing systems has arisen to make sense of data \cite{Shao12}. 

In addition to handling large volumes of data, organisations need to make faster decisions (e.g. early detection of fraud in online auctions \cite{Eberle07}; online recommendations or click stream processing \cite{Ananthanarayanan13}). Unfortunately, this task has become increasingly difficult given that we produce information much faster than ever before. There are few graph analysis systems that can deal with large-scale graphs in near real-time\footnote{\tiny Near real-time is highly dependent on the application. In this manuscript, near real-time indicates that responses are needed faster than the time it takes to distribute the graph files to all machines in the cluster, split the graph across machines and load the graph in memory.}. 

To make matter worse graphs are changing entities (like the real-world systems they represent) \cite{Holme12,Gaito2012,Leskovec2005}. As we will see below, only a handful of systems are capable of responding in near real-time at scale, while adapting to changes in:  1) the information stored in edges and nodes (e.g. people's age, router load); 2) the graph topology (e.g. adding new users/edges or considering user churn) \cite{Klein94,Henzinger1999,Leskovec2005,Ediger12}; 3) the type of graph processing or load required by end users. For example, banks must be able to detect frauds as soon as possible \cite{Weigert2011}, and analyse call records in telecom networks \cite{Cortes03,Eberle07} and news article recommendations in social networks must be delivered within minutes \cite{Cai12}. 

Coping with many changes in large-scale graphs while preserving near real-time processing times is a challenging task. A broad perspective on the recent advances is missing in current literature review efforts.

In this paper we review the main strategies to create large-scale near real-time graph processing systems and provide readers with the most salient examples (Section \ref{sec:gralSOTA}). In Section \ref{sec:dynamic}, we present this need for dynamic adaptation and the types of runtime changes that can affect a graph processing system. Adaptation comes at a price: additional overhead and performance degradation. The techniques used by some systems to prevent this performance decay are presented in Section \ref{sec:dynsys}. In Section \ref{sec:discussion} we discuss the main features and current trends in large-scale near real-time graph processing. Finally, Section \ref{sec:conclusion} wraps up the main conclusions of this analysis.


\section{Divide and Conquer}
\label{sec:gralSOTA}

The most straightforward way of processing a graph is loading the graph in the memory of a single machine. Large graphs can be handled in a standard desktop machine by accessing graph data from the hard drive. The main memory only holds the graph elements needed for a computation. This approach is followed by graph databases like Neo4j\footnote{\tiny Neo4j provides no built-in support to scale to multiple machines, although your application must create its own sharding layer} \cite{Webber12}, see top-left quadrant in Table \ref{table:classification}. Disk access times are slow; with smarter (not random) access to disk data resulting in significant performance improvements processing large-scale graphs \cite{Kyrola12}. However, disk-based systems are not the best option for large-scale scenarios because throughput is limited by the lack of parallelism and near real-time processing cannot be achieved due to large disk seek latencies.

There are several single machine systems that keep the whole graph in memory to help deliver timely responses and avoid unnecessary usage of the disk (used just to load and persist data) \cite{Bitsy13,Jstor13,Jung13,igraph13}. However, these systems do not scale to large graphs, being unable to host the whole graph in memory. The graph must, therefore, be partitioned into separate subgraphs to be stored and processed in different machines.

Graph-oriented distributed databases like Titan, FlockDB or AllegroDB \cite{Titan13,Flock13,Allegro13} store massive amounts of information on disks to enable a variety of graph queries (see top-right cell in Table \ref{table:classification}). These systems can be adapted to support graph dynamism (adapting the topology of time dependent graphs, see \cite{Cattuto12}, for instance). Unfortunately, disk-based access to data makes computation too slow for near real-time large-scale graph processing.

From the discussion above it seems that we need horizontally scalable distributed systems and in-memory processing to cope with near real-time large scale graphs. The need for fast processing at scale has led to the creation of a myriad of distributed systems that load the whole graph in memory \cite{Najork09, Malewicz10, Cheng12, Low12, Zeng12, Prabhakaran12, Mondal12, Yang12, Hama13,Giraph13,Faunus13, Khayyat13, salihoglu13, Ugander13, Vaquero13, Jouili13}. 

\begin{table*}
  \scriptsize
  \begin{center}
    \begin{tabular}{| l || c | c |}
    \hline
     & centralised & distributed \\ \hline \hline
    disk & databases (no sharding) \cite{Webber12} & \cite{Titan13,Flock13,Allegro13,Low12,Faunus13}  \\ \hline
    memory & \cite{Bitsy13,Jstor13,Jung13,igraph13} & \cite{Najork09, Malewicz10, Cheng12, Zeng12, Prabhakaran12, Mondal12, Yang12, Hama13, Giraph13,  Khayyat13, salihoglu13, Ugander13, Vaquero13, Jouili13} \\
    \hline
    \end{tabular}
  \end{center}
  \label{table:classification}
  \vspace{-15pt}
  \caption{Broad Classification of Large-Scale Graph Processing Sytems.}
\end{table*}

\subsection{Partition Optimisation}

The advantages of distributed nature of these distributed in-memory systems come with a price to be paid. Partitioning the graph unavoidably results in additional communication overhead between machines and the risk that the workload might be unbalanced across the machines (uneven number of vertices/edges). 

Most distributed systems partition the graph using a hash function (typically the integer id of the vertex modulo \# of partitions) that renders balanced partitions at no cost in terms of keeping a ``directory'' to locate vertices. While highly scalable, this approach is pseudo-random and leads to neighbouring vertices frequently being assigned to different machines \cite{Malewicz10, Stanton12, Vaquero13}. 

With graph partitioning playing a fundamental role in distributed systems performance,  location randomness makes communication overhead worse than it would be with an exhaustive partitioning strategy. Finding optimal partitions is an NP-hard problem \cite{Meyerhenke09} and many heuristics have been developed to reduce random partitioning communication overhead and create balanced partitions (see \cite{Ugander13,salihoglu13,Stanton12, Yang12, Mondal12} for recent examples). 

Unfortunately, smart partitioning has just been applied right at the time the graph is initially loaded in memory. For instance, \cite{Khayyat13a} propose that detecting whether or not the topology of the graph follows a power law distribution in the vertex degree may be a nice hint to apply different processing engines aimed to reduce communication overhead. \cite{Khayyat13a} relies on ParMETIS \cite{parmetis} for the initial partitioning of the graph (unlike most other large scale distributed system, which rely on static hash-based partitioning).

Stanton and Kliot \cite{Stanton12} use a streaming heuristic to place vertices close to where most of their previously seen neighbours have been allocated (accumulating a partial global view of the graph). These systems cope well with static graphs where neither the workload nor the topology changes at runtime (streaming the whole graph would be too expensive, see paragraph on dynamic partitioning in Section \ref{sec:dynsys}). 

Other systems perform similar load time optimisation by placing neighbouring vertices together in an iterative manner \cite{salihoglu13}, but do not enable the topology to change in real-time. 

Next section provides an overview of the dynamism level supported by different large-scale processing systems and how near real-time features are typically supported.

\section{Dynamic Graphs}
\label{sec:dynamic}

Graphs change as often as the real-world entities and relationships they represent. Change can be fast and bursty (see \cite{Gaito2012}), which brings new challenges to the systems processing the graph to obtain results. Changes during graph processing can roughly be categorised as:

\begin{itemize}

\item \textbf{The number and types of queries} (workload) on the graph may change responding to random or seasonal peaks. As graphs are partitioned over several machines, changes in the workload can create hotspots that impact performance of the system. A common approach consists on dynamically assigning parts of the graph to different machines. Dynamic load balancing strategies can then be employed to make a more efficient usage of the available resources and reduce query time.

\item \textbf{The information contained in the graph} (e.g. apps a Facebook user is currently using) is continuously changing. In order to improve the accuracy of the responses to our queries, it would be desirable to operate with current information only. The change in information will render  previous computations on the graph obsolete, thus forcing systems to provide update information on a near real-time basis.

\item \textbf{The vertices} (e.g. users who logged in during the last week/month) or \textbf{edges} (e.g. calls connecting people during the last couple of hours) that are modelled in the graph can change, altering the graph topology over time. The speed at which these changes occur will depend on the information being modelled (for instance, an incoming call can trigger the creation of an edge and new vertices).  Changes in the graph topology impact computation performance in two ways: it can affect load balancing between machines; and it can also affect partitioning, and therefore create communication overhead.
\end{itemize}

These changes impact graph processing: Dealing with obsolete information or using a partitioning that creates hotspots on a few partitions, keeping the rest idle, is not what most applications require. Therefore, systems need to be able to accommodate a variety of changes in the graph at runtime.

\subsection{The Curse of Dynamism}
\label{sec:curse}

Modifying the information contained in the graph at runtime does not present significant technical challenges for current systems. On the other hand, modifying the topology of the graph while some computations are performed and avoiding overloaded partitions and hotspots is harder to achieve. 



A recent generation of systems has been designed to support runtime changes to the topology of the graph \cite{Cheng12,Ediger12,Shao13}. Unfortunately, these systems do not perform any optimisation when the structure of the graph changes. This lack of adaptation to changes in the topology of the graph conveys a gradual degradation of the partitioning of the graph (more neighbours are placed in a separate partition/machine) that negatively affects performance. 

\begin{figure}[h!]
  \centering    
      \includegraphics[width=0.5\textwidth]{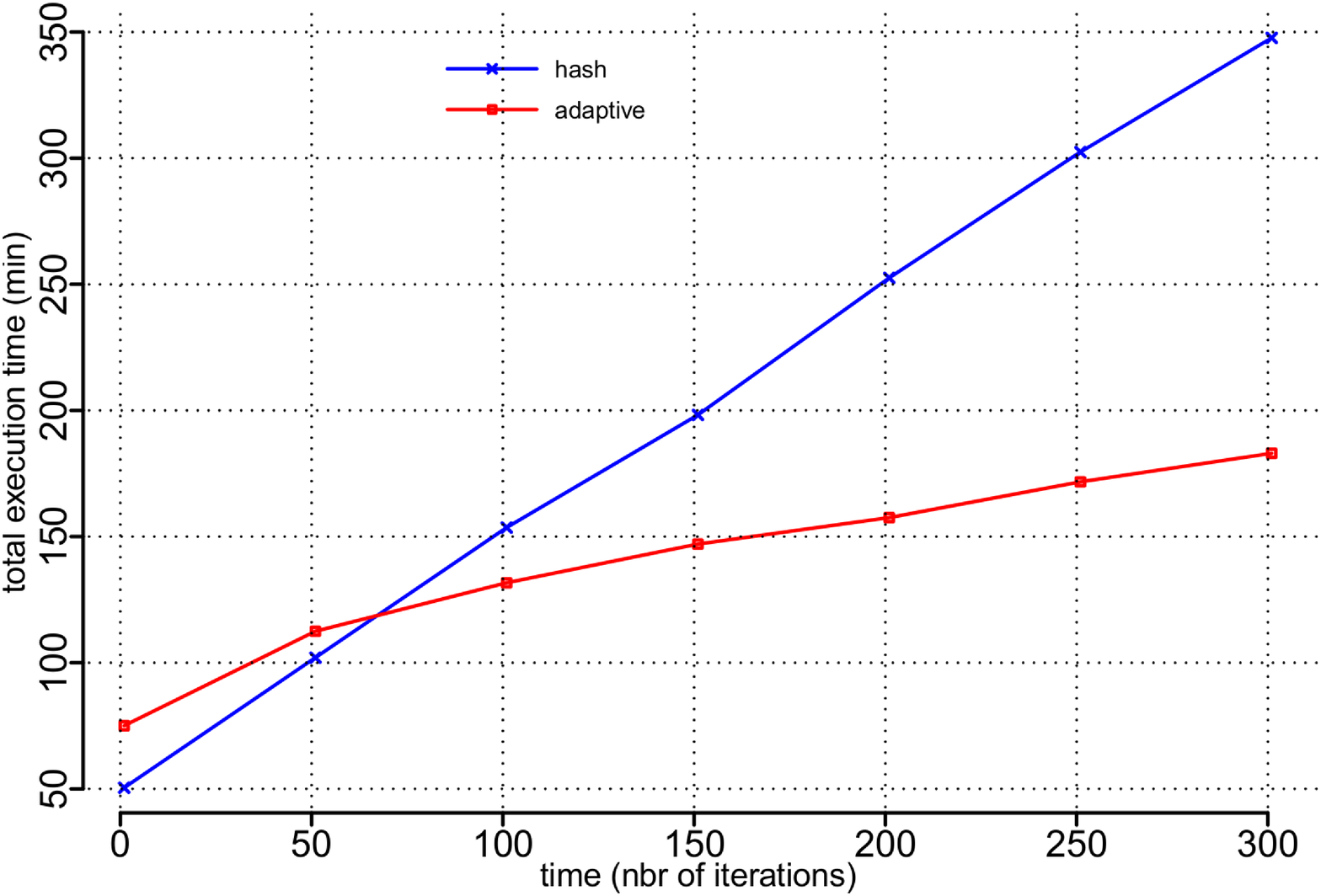}
  \vspace{-20pt}
  \caption{Comparison of static hash  and  adaptive partitioning strategies.}
 \label{fig:obsolete}
\end{figure}

Figure \ref{fig:obsolete} shows the effects of adding new edges and vertices in a Pregel-like system. Following the Bulk Synchronous Parallel (BSP) model, separate processors compute some operations separately and then exchange messages to sync up on a global barrier. The process goes on on a series of iterations until no more messages are exchanged. In this figure, the partitioning of the vertices across machines can either be static (hash) or dynamic (using a modified label propagation heuristic) to increase vertex locality and reduce communication overhead across machines. As can be observed, the performance (execution time) of \textit{de facto} standard partitioning technique (hash) gets gradually worse as new vertices are added on every iteration, while an adaptive strategy performs better \cite{Vaquero13}. This degradation (induced by loss of data locality) can be seen as a ``curse'' for large scale graph processing systems. 

\section{Dynamism in Large-Scale Graph Processing Systems}
\label{sec:dynsys}

This section presents the most widely used techniques for runtime adaptation to overloaded partitions and hotspots that can appear under changes in the workload or the topology of the graph. 



Some systems support dynamic optimisations of runtime partitioning \cite{Mondal12,Yang12,Vaquero13} (see Table \ref{table:dynSys}) and some Internet giants use similar techniques in their graph processing engines (see \cite{Ching13}). The few systems that support dynamic optimisation of their partitioning at runtime present two differentiated techniques to support this dynamism:

\begin{table*}
  \scriptsize
  \begin{center}
    \begin{tabular}{| c | c | c | c | c |}
    \hline
    \textbf{Reference} & \textbf{Variable workload} & \textbf{Variable topology} & \textbf{Load time optimisation} & \textbf{Runtime optimisation}   \\  \hline
     \cite{Stanton12} & No & No & Yes & No \\ \hline
     \cite{salihoglu13}& No & No & Yes & Yes \\ \hline      
     \cite{Khayyat13} & Detecting graph type & No & Yes & No \\ \hline
     \cite{Ugander13} & No & Yes & No & No \\ \hline
     \cite{Jouili13} & No & Yes & No\footnote{\tiny Virtual edges reduce the number of messages across machine without altering partitioning.} & No \\ \hline
     \cite{Cheng12} & No & Yes & No & No\\ \hline
     \cite{Shao13} & No & Yes & No & No \\ \hline 
     \cite{Ching13}& No & No & No & Yes \\ \hline
     \cite{Mondal12} & Dynamic replication & Yes & Yes & Yes \\ \hline
     \cite{Yang12} & Dynamic replication & Yes & Yes & Yes \\ \hline
     \cite{Vaquero13}& No & Yes & Yes & Yes \\ \hline
    \end{tabular}
  \end{center}
  \label{table:dynSys}
  \vspace{-15pt}
  \caption{Degree of Dynamism Exhibited by Different Large-Scale Graph Processing Systems.}
\end{table*}

\begin{itemize}

\item \textbf{Vertex Migration} Cut edge minimisation based on ``getting neighbouring vertices into the same partition'' (a form of label propagation and clustering) run on an iterative system (like Pregel, see description of the BSP model above) every $N$ iterations, where $N$ varies on how fast the graph changes \cite{Vaquero13}. 

\item Hotspots (heavily-used \textbf{Partitions} ) \textbf{replication} in less loaded machines and load balancing is a classic technique to adapt to workload variations. For instance, incremental optimisation techniques for adapting the way graphs are distributed (partitioned) across processors depending on workload are well established \cite{Walshaw97}. These techniques have been introduced in a few dynamic systems \cite{Mondal12,Yang12}.

\end{itemize}

These two techniques can also be labelled as ``dynamic partitioning'' and ``dynamic replication'', respectively. Table \ref{table:dynTopo} highlights the most important differences between them. Dynamic partitioning focuses on delivering high throughput for a reduced number of users that constantly execute the same query (or a reduced number of queries) over a large portion of the graph. Dynamic replication performs best when low latency responses are required for a large number of queries that span a very limited region of the graph (neibourghood of a node, or ego network). 

Both techniques present limitations of their own: dynamic partitioning systems \cite{Vaquero13,Ching13} may incur in vertex migration overhead, while dynamic replication systems \cite{Mondal12,Yang12} have to live with the cost of dynamically creating new partitions. While lazy-replication techniques \cite{Mondal12} or lazy migration techniques (do not migrate a vertex unless it is active) can definitely alleviate this overhead, the rate of change in the topology and the workload at runtime are still bound by the cost of replication/migration and the available resources.

\begin{table}
  \scriptsize
  \begin{center}
    \begin{tabular}{| c | c | c |}
    \hline
     & \textbf{Dynamic partitioning} & \textbf{Dynamic replication}  \\  \hline
    \textit{output} & high throughput & low latency  \\ \hline
    \textit{query type} & continuous/whole graph & short/ego networks  \\ \hline
    \textit{\#  of users} & few & many \\ \hline
    \textit{failure-tolerance} & snapshotting & existence of replicas  \\ \hline
    \end{tabular}
  \end{center}
  \label{table:dynTopo}
  \vspace{-15pt}
  \caption{Differences between the Two Main Techniques to Introduce Runtime Optimisation in Large-Scale Graph Processing Systems.}
\end{table}

\subsection{Dynamic Partitioning}
Dynamic partitioning algorithms have been there for a while in the world of mesh networks in parallel scientific computing (see \cite{Catalyurek2009,Schloegel2001,fourestier2013} for recent review on the topic). According to this community, dynamic partitioning methods can broadly be classified in: 1) scratch-map: methods that create a new set of partitions from scratch; 2) incremental: methods that iteratively update existing partitions in order to minimise the cost of migration, the communication cost or both.

\paragraph{Scratch-map}
Reloading the graph from scratch is an option that can initially be considered (one could easily take stream partitioning techniques such as \cite{Stanton12} and execute them with some periodicity). More sophisticated restreaming techniques have recently been introduced to help reload the graph (assuming not many changes have occurred) \cite{Nishimura2013}. Nishimura and Ugander show how to restream state of the art stream partitioning methods. These re-streaming mechanism make a new pass of the graph, which may not scale even when partition parallelisation is doable in separate workers (streaming the whole graphs is still required on every parallel worker). As an illustrative example, partitioning a billion edge/vertex scale graph may take $\simeq$ 1.4h \cite{Tian13}. Also, there is a high volume of communication between workers between restreams, each worker reports on their share of the partitioning and this compiled list is distributed to all workers for the next restream. Restreaming and its 
synchronisation needs would reduce the potential for real-time processing of the graph.

\paragraph{Incremental Methods}
ParMETIS \cite{parmetis} is arguably one of the most used diffusive iterative techniques. It leverages parallel processing for partitioning the graph, through multilevel k-way partitioning, adaptive re-partitioning, and parallel multi-constrained partitioning schemes. While its hierarchical approach is excellent for FEM networks, it requires global visibility during the initial partitioning phase: all the pieces of the graph are scattered to all threads using an all-to-all broadcast operation. Then, each process explores a single path of the recursive bisection tree. Indeed, ParMETIS relies on Fiduccia-Mattheyses algorithm, which makes heavy use of some sort of bucket data structure to sort all possible vertex moves per descending edge cut gain. Upon migration, candidate vertices are taken out of the bucket and the algorithm checks whether moving them would preserve balanced partitions. If not, the vertex is put aside and the next one is taken out of the bucket. Once a vertex has been moved all the vertices 
kept aside are reintroduced in the bucket. While this works well for a single partitioning, it is extremely ineffective when applied to repartitioning. If migration costs are high, the algorithm will move as few vertices as possible. If the graph has changed much, the algorithm has to move enough vertices to keep balanced partitions. The best partitions are not balanced ones, therefore all the vertices that should be moved to minimise edge cuts cannot be accepted because they will make imbalance even worse.

While scratch-map methods are poorly scalable and render bad results when the adaptation is light or is scattered trough the whole graph, incremental and repartitioning methods can bring globally non optimal solutions when topological changes are important in a localised area of the graph. 

Most of the available general-purpose graph processing systems have not relied on the wisdom of dynamic mesh partitioning (except for \cite{Khayyat13,Vaquero13}), being hash the prevalent approach to partition. This is due to the fact that some of these techniques are difficult to parallelise and implement in a distributed system \cite{Tian13}. Also, \cite{Walshaw95} indicate that adaptive meshes refer to a slightly different problem: that where the evolution of the computation causes a variation in the position and density of datapoints in the mesh, forcing the repartitioning process to balance the load. The addition of new data points is restricted to the static structure of the mesh. Beyond scientific computation, many networks present a preferential attachment model. More general solutions are needed that do not rely on predefined static structures so that new edges/vertices can be added anywhere.

\subsection{Dynamic Replication}

Having a partitioned graph that exploits locality can result in hotspots when queries are confined to a small region of the graph, as queries are constrained to a neighbourhood of interest. Replication strategies dynamically detect and create replicas of the hotspots as needed to load balance between them, while keeping very low latency in the responses offered to queries. Replication can be done at different levels of granularity: 1) Node level (see \cite{Pujol2010, Khan2011, Sun2012}); 2) Partition level (see \cite{Yang12}); 3) Clusters within partitions (k-means clustering of the nodes in a partition is a classic approach to this problem, see \cite{Mondal12}); 4) Graph level. 

While node replication may work well for small graphs, the number of node replica updates needed to maintain local semantics (any query can be completed locally so that no external pulls are needed) is prohibitive when the size of the graph increases. Full graph replication is only doable for small enough graphs. Therefore, partition/cluster level replication are the preferred approaches for large graphs. 

A static partition can be created and new replicas created as hotspots are detected like in \cite{Mondal12}. Also, Self Evolving Distributed Graph Environment (Sedge) and similar approaches \cite{Yang12} create a fixed set of initial partitions and replicate whole partitions or create new ones based on the types of queries being executed. These two levels of partitioning (primary/static and secondary/query-dependent) enable Sedge to perform queries efficiently upon changes on the topology of the graph, since new secondary partitions will be created. Unfortunately, keeping primary partitions static in a rapidly changing graph may rapidly worsen performance.

The key information for making replication decisions are read/write access patterns for different nodes. For every cluster/partition the access patterns of its nodes are aggregated and decisions are made to place a new replica on a less overloaded machine. The timing for creating the replicas is very application-dependent, as a tradeoff between responsiveness (e.g. react to flash traffic) and minimising replication costs needs to be achieved depending on typical workload traces of the application.

Replication is naturally coupled  with load balancing strategies. Mondal and Pal categorise dynamic distributed load balancing strategies depending on their distribution, cooperativity, or heuristic nature \cite{mandal2010}; see also \cite{Willebeek-LeMair93,Teo01comparisonof} for review.

\paragraph{Choosing the Right Approach}

\begin{figure}[h!]
  \centering    
      \includegraphics[width=0.3\textwidth]{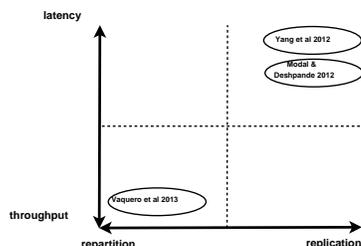}
      \vspace{-15pt}
  \caption{Categorisation of Systems Depending on the Runtime Optimisation Technique.}
 \label{fig:sota}
\end{figure}

Figure \ref{fig:sota} tries to sum up which approach is being used by each of the systems supporting dynamic optimisation of their partitions.

One way to classify workloads is to differentiate on whether to answer a query all nodes/edges need to be touched (e.g., to estimate diameter, compute PageRank) or only a small subset (e.g., two-hop neighbours of a node). Different optimisation techniques are good for each. 

\section{Discussion}
\label{sec:discussion}

\paragraph{Scale of the Scalability}

We have been discussing large-scale systems without actually offering any figures on any scale metric. Table \ref{table:scale} offers an overview of scalability\footnote{\tiny Figures in this table should be taken as a reported maximum value, not an actual capping on the capabilities of the systems. It is very likely that, given their scalable and distributed architectures, most of these systems could reach higher given a larger cluster. Handy datasets used by some of these work are available at http://law.di.unimi.it/datasets.php and http://snap.stanford.edu/snap/}. As can be observed, most systems do not process graphs beyond tens of millions of vertices and there is a handful of them that routinely process 1 billion node graphs (e.g. \cite{Shao13,Ching13}), but the amount of information processed (RAM used) by the graph is similar to other systems (a few tens of GB). 

Graph algorithms usually  traverse parts of the graph and are therefore inherently iterative (one would write a for loop for all the unvisited nodes, for instance). However, there is a surprising lack of reported values for metrics on how much data the system moves around per iteration or how many messages are sent. The overall number of messages and the way vertices are activated to send/receive messages may play a key role in performance, as shown by \cite{redekopp14}.

\begin{table*}
  \scriptsize
  \begin{center}
    \begin{tabular}{| c | c | c | c | c | c |}
    \hline
    \textbf{Reference} & \textbf{\# vertices} & \textbf{\# edges} & \textbf{Used RAM} & \textbf{\# msgs/iteration} & \textbf{Data transfer per iteration/replication}     \\  \hline
     \cite{salihoglu13}&100M & 3.7T & 14 GB & NA & NA \\ \hline      
     \cite{Khayyat13} & 17M & 70M & $<2$ GB & NA & 16GB \\ \hline
     \cite{Ugander13} & NA & NA & NA & NA & NA  \\ \hline
     \cite{Jouili13} & 4.7M & 70M & < 75 GB & NA & NA  \\ \hline
     \cite{Cheng12} & 8M & 29M & $<512$ GB & 100000 & NA\\ \hline
     \cite{Shao13} & 21G & 160G & 1.46 TB & NA & NA \\ \hline 
     \cite{Mondal12} & 1.8M & 18M & $<12$ GB & $<8$ M & NA \\ \hline
     \cite{Yang12} & 40M & 1.4T & 14GB & NA & NA \\ \hline
     \cite{Vaquero13}& 100M & 600M & 3TB & 1.2G & 0.3TB \\ \hline
    \end{tabular}
  \end{center}
  \label{table:scale}
  \vspace{-15pt}
  \caption{Scale Level Reached by Different Large-Scale Graph Processing Systems. NA means not available (in BSP-based systems this implies the time per iteration has not been reported, while in replicated systems it means the runtime model may not be iterative as queries can be resolved on a local partition at once).}
\end{table*}

\paragraph{Near Real-Time, Really?}

A second crucial aspect of this work is near real-time potential. In a distributed system, the number of messages and size of the data moved around per iteration/replication is expected to be large. This would, in turn, increase the response time of the system. Some applications, like finding auctions matching given words in a search engine, require ms responses \cite{Ananthanarayanan13}; other applications, such as call record analysis, are reported to run on a weekly basis \cite{Cortes03}, but would benefit from  faster turnaround. 

\begin{table}
  \scriptsize
  \begin{center}
    \begin{tabular}{| c | c |}
    \hline
    \textbf{Reference} & \textbf{time per iteration}    \\  \hline
     \cite{salihoglu13}& 1.5 min  \\ \hline      
     \cite{Khayyat13} & 4 min \\ \hline
     \cite{Cheng12} &  NA \\ \hline
     \cite{Shao13} &  NA \\ \hline 
     \cite{Ching13}& 4 min \\ \hline
     \cite{Vaquero13}& 1.5 min \\ \hline
    \end{tabular}
  \end{center}
  \label{table:rt}
  \vspace{-15pt}
  \caption{Minimum Time per Iteration Reported by Different Large-Scale Graph Processing Systems. NA means not available.}
\end{table}

\begin{table}
  \scriptsize
  \begin{center}
    \begin{tabular}{| c | c |}
    \hline
    \textbf{Reference} & \textbf{time per query}    \\  \hline
     \cite{Ugander13} &  50 ms   \\ \hline
     \cite{Mondal12} & NA  \\ \hline
     \cite{Yang12} &  2.5 ms\\ \hline
    \end{tabular}
  \end{center}
  \label{table:rt2}
  \vspace{-15pt}
  \caption{Minimum Time per Query Reported by Different Large-Scale Graph Processing Systems. NA means not available.}
\end{table}

We collect in Tables \ref{table:rt} and \ref{table:rt2} the time per step reported by these iterative systems. As can be observed iteration times are at least a few minutes, since their queries typically involve traversing a large portion of the graph. Replicated systems can resolve several hundreds of queries per second benefiting from locality of the data in each partition an load balancing across replicas. These figures hold true for dynamic systems, implying that the vertex migration/replication process is optimised for the problems at hand in order to reduce the overhead imposed by the re-optimisation process. Current research is aimed at reducing iteration/query times, making graph analysis systems as fast as possible.

\paragraph{Planned Obsolescence}

In addition to large scale and near real-time, dynamism was the third feature we have surveyed. In the context of dynamically changing topologies graph processing systems have to decide on two main aspects: how does the graph topology naturally evolve over time, and when and how do we reflect these changes in the graph processing system (the way vertices/edges are added/removed from the runtime engine). There are several elements that deserve careful consideration to avoid generating ghost topologies that exist for a very limited time, but alter the results or sticky topologies that prevent us from seeing what is actually changing: 1) Definition of active vertex/edge; 2) Timing for adding active vertices/edges  to the runtime engine; 3) Definition of inactive vertex/edge; 4) Time for deleting inactive vertices/edges from the runtime engine; 5) Order of addition/deletion of vertices/edges (assuming changes are buffered and applied all at once); 6) Burstiness of the changes in the graph \cite{Gaito2012}.

\begin{table*}
  \scriptsize
  \begin{center}
    \begin{tabular}{| c | c | c |}
    \hline
    \textbf{Reference} & \textbf{Addition}  & \textbf{Deletion}  \\  \hline
     \cite{Cortes03} & Change buffering & Change buffering\\ \hline
     \cite{Ugander13} & NA & NA \\ \hline
     \cite{Cheng12} & Change buffering & Does not delete \\ \hline
     \cite{Shao13} & NA & NA\\ \hline
     \cite{Mondal12} & NA & NA \\ \hline
     \cite{Yang12} & NA & NA\\ \hline
     \cite{Gaito2012} & Bursts of vertex/edge addition & NA\\ \hline
     \cite{Vaquero13} & Preferential attachment and change buffering & Random deletion, preferential deletion and change buffering \\ \hline
    \end{tabular}
  \end{center}
  \label{table:adddel}
  \vspace{-15pt}
  \caption{Methods Employed to Add/Delete Vertices and Edges to/from the Running Graph. NA means not available. The high number of NAs in this table is indicative of how little work has actually been done in this space.}
\end{table*}

Cortes et al. \cite{Cortes03} explain the importance of graph topology changes and present a sliding window criteria for adding and removing elements based on an incoming stream of information. They show that new nodes (nodes seen in week $i$ that were not there in week $i-1$ ) are added and old ones (nodes not seen at week $i$ that were seen at week $i-1$) deleted at a similar rate of approximately 1 \% per week. While nodes in their user network were very stable, edges seemed to be more volatile having higher replacement rates.  Cheng et al. \cite{Cheng12} provide a configurable regular interval in which they buffer new tweets before applying changes to the graph. Window size is important for timeliness (availability of the changes for processing). Cheng et al. and Vaquero et al. \cite{Cheng12,Vaquero13} were the only ones to provide some figures on timeliness related to the number of new nodes to be added. More detailed studies are needed to draw any definitive conclusion.

Many models of how real world graphs expand report a preferential attachment model \cite{Leskovec2005} and apply to real-world phenomena (e.g. death of myocytes during cardiac infarction would imply preferential deletion of ``nodes'' in the vicinity of the infarcted area \cite{Vaquero13}). 

Table 6 summarises the main techniques (buffering and preferential addition/attrition) observed in dynamic graphs. More experimental analysis on dynamic graphs are required to determine and understand the processes ruling topology changes. In general, dealing with highly dynamic graphs or graphs where bursts of changes occur (see \cite{Gaito2012,Vaquero13}) calls for continuous optimisation that can rapidly absorb changes, while sparse optimisation runs may be enough for slowly changing graphs (\cite{Ugander13}).

Another important aspect from the systems perspective is how to efficiently apply these changes to the topology. Cheng et al. study the effects of adding more of these change-applying entities (parallelising changes) and reveal how throughput can be increased by parallelisation.

\paragraph{Can We Have It All?}

The analysis of the available art has revealed how complicated it is to achieve near real-time responses at scale; much more so in a dynamic context. Distributed systems classically reach near real-time capabilities at large scales by relaxing the accuracy of the obtained results. The idea is that one can live with less precise responses as long as they are delivered fast (see \cite{DasSarma08} or BlinkDB \cite{Agarwal13} as recent examples).


Indeed, graph processing systems such as Trinity \cite{Shao13} also perform some tradeoffs between accuracy and real-time/scale: the authors claim that they ``perform graph computation locally on each machine and then aggregate their answers to derive the answer for the entire graph, or can we use probabilistic inference to derive the answer for the entire graph from the answer on a single machine''.

\begin{figure}[h!]
  \centering    
      \includegraphics[width=0.3\textwidth]{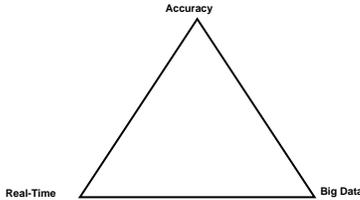}
      \vspace{-15pt}
  \caption{Big data, Accuracy and Real-time (BAR) Conjecture}
 \label{fig:bar}
\end{figure}

These three elements (large-scale, accuracy and real-time) can be visualised as a triangle (see Figure \ref{fig:bar}). Most systems strive to contain all three vertices of the triangle, but graph analysis systems, like other distributed systems, typically cope with just two of the variables in the triangle.

\paragraph{Trends in System Architectures}

GPU-enabled graph analytics systems vastly boost the available compute power \cite{Harish07}. It is just recently that they have reached a very mature status for large-scale graph analytics \cite{gharaibeh13}. While these architectures are very likely to gain more relevance in the short term, they do not solve some of the main problems for near real-time responses we found in our analysis: distribution and disk access make systems incur in higher than acceptable latencies. 

Malewicz et al. \cite{Malewicz10} report how the message passing rate and the volume of information exchanged between worker nodes causes link congestion. Available RAM is also a limiting factor. Larger memory banks would allow for larger partitions and reduced overhead. This is worsened in very dynamic scenarios where vertices need to be moved or replicated. 

Industrial efforts follow common practise in the research community, like keeping everything in memory, and advocate for more centralised solutions where a single machine hosts all the data/processing in massive memory banks (getting rid of network congestion and disk seeks)\footnote{\tiny http://hpphenom.blogspot.co.uk/2013/05/project\-kraken.html\\http://www54.sap.com/pc/tech/in-memory-computing-hana/software/overview/index.html}. The gradual introduction of Solid State Drives in the mass market also highlights this trend. As prize and latency of memory decrease and its capacity increases we are likely to see how most systems go for single machine in-memory designs (see Table \ref{table:classification}). 

In this setting of hundreds of cores using shared memory in a single machine, new techniques for concurrency control and cache coherence may take over current distributed system designs. These centralised massive memory banks face the challenge of failure tolerance management, where properly implemented distributed systems excel. More research is granted in this space too. 

Parallel Random Access Machine (PRAM) systems keep in globally-shared memory and there is
implicit communication by updating memory. Some PRAM systems have specialised architectures for graph processing, like Cray XMT2, BlueGene or SGI UV2. Cray XMT2, for instance, tolerates high memory latencies using massive hardware multithreading. Fine-grained synchronization constructs are supported through full-empty bits and atomic fetch-and-add instructions. Each processor within a Cray XMT contains 128 hardware streams that may block temporarily while waiting for a long-latency instruction to return. The processor will execute one instruction per cycle from hardware streams that have instructions ready to execute. 

This survey has focused on distributed memory systems where data is distributed to local memory partitions and communication takes place by explicitly sending messages across them for the following reasons: 1) they run on commodity machines and scale by adding more machines (synchronization, deadlock, hot-spotting, and others can be barriers to obtaining linear scalability in shared-memory systems \cite{Ediger2013}); 2) they can be ported to a cloud, reducing operational costs (at 10 cents per KWh, this implies \$7 million per year to keep lights on\footnote{\tiny Energy specs from: http://www.top500.org/list/2012/11}); 3) implementing parallel graph algorithms in large, shared memory machines, such as the Cray XMT, can be challenging for programmers \cite{Ediger2013}.

Most graph processing algorithms force systems to do lots of random access to memory causing many cache and TLB misses. Same way an organised access to disk enabled systems in a single machine to scale \cite{Kyrola12}, these techniques could be applied to make a smarter use of the memory hierarchy of the machines in the cluster.

\paragraph{Algorithms for near Real-time Large-scale Dynamic Graph Processing}

Algorithms should ideally be independent of the system executing them. The actual implementation of these algorithms is also determined by the architecture of the underlying system. In this section we analyse current support for dynamic graphs from a more algorithmic perspective.

Faloutsos and Kang \cite{Faloutsos12} describe algorithms for mining large-scale graphs. The algorithms they deal with are general and cover diverse cases including structural analysis, eigensolver, storage/indexing, and compression. Unfortunately, no attention is paid to real-time processing and the effect of runtime optimisation on the accuracy of the obtained results.

The idea of building dynamic graph algorithms that could update the result of the computation in near real-time is not at all new. Cooke and Halsey \cite{Cooke66} pioneered this area back in the 60s. More recently Klein et al. \cite{Klein94} continued this work by enabling dynamism for shortest path graph algorithms. Henzinger and King \cite{Henzinger1999} presented a set of fully dynamic algorithms that maintain connectivity, bipartiteness, and approximate minimum spanning trees in poly-logarithmic time per edge insertion or deletion. 
 
Following this line of work, most of the algorithms that support dynamic changes are focused on introducing a temporal dimension on classic topological metrics of the graph like time-respecting paths, connectivity, distances/latencies, network efficiency, centrality, patterns/motifs, etc. (see \cite{Holme12} for recent review). Let us  give a few recent examples for illustration purposes only. Nicosia et al. \cite{Nicosia12} re-define the concept of connectedness to consider nodes that are connected by a path that is never fully there, but can be built in different time stages. Similar work has arisen from the field of delay-tolerant networks were paths are re-defined as “journeys” \cite{Santoro11}. Journeys can be seen as a particular kind of path whose edges do not necessarily follow one another instantly, but instead induces waiting times at intermediate nodes. As a consequence all the concepts built on top of paths take a temporal dimension themselves \cite{Santoro11}. This idea has also been applied to detecting overlapping time dependent communities, creation of clusters or analysis of the dynamism of disconnected components \cite{Palla07,Berlingerio10,McGlohon08,Zheleva09}.


Some work has been done beyond topological measurements: information-theoretical adaptive algorithms. \cite{timo2005} use network entropy for detecting temporal uncertainty in communication networks. Also, some studies focus on measuring vertex/edge dynamism and hiw changes occur (see \cite{Holme12}). More work is needed to make dynamic other algorithms for graph analysis that are not purely focused on the topology of the graph, but also on the data associated to each element in the graph. This is a challenging task. When the information of the graph or the topology of the graph itself changes before the computation has finished, the algorithm may return inaccurate results. This may be reasonable in some cases (slight variations in Page Rank get eventually fixed in an iterative processing of the graph), but it may have catastrophic consequences for some other applications (e.g. a graph representing the stock exchange in a High Frequency Trading setting). 

The algorithmic examples given so far are based on incremental approaches to update the topology of the graph. Approximation strategies to graph analysis have also been around for a while and new solutions keep appearing \cite{DasSarma08,Agarwal13}. The use of these approximations will most likely simplify the design of far more scalable and nearer real time systems (as hypothesised by the BAR conjecture above).


Even when the body of work in graph algorithms that can adapt to changes in the graph at runtime is still small, the relevance of this type of problems will likely bring more research and new solutions to this area in the near future.

\paragraph{Other Aspects}

While this short survey has tried to focus on the large scale near real-time dynamic graphs, there is a myriad of other features that should be considered in the system. While it is beyond the scope of this survey to have an exhaustive list of them, we felt obliged to mention a few important issues that will gain relevance in the short term.

Most of the graph processing systems above support a single algorithm to be run at one time by a single user (multi-tenancy is not easily supported at this scale). Some recent attemps try to let several users share the same data by creating ``job-specific'' vertices \cite{Yang2013}. Unfortunately, all thevertices execute the same algorithm. More generic software platforms that support the concurrent execution of several algorithms  graph processing are needed. 

A second aspect of the usability is not just multitenancy, but also the ease of use (complexity) of the programming model. Many of the systems above are based on Google's Pregel. Still most business analysts like to think of problems in terms of SQL queries on a set of entities in their data warehouse. Also, scientists tend to analyse their data using popular statistical packages like SPSS or R or mathematical libraries like Mathematica or Matlab. Ideally one would like to treat graphs as a first class data structure (no need to traverse it from the analyst's perspective) that has a set of associated functions (e.g. \textit{pagerank(graph)}). Therefore, it seems that functional \cite{Oliveira2012}, data flow \cite{Tran2012} and domain specific \cite{Hong2012} graph analysis languages are being created to ease our graph processing tasks.

Beyond the user's view of these graphs, running these very large graphs implies having very large underlying systems. Data centre operators are also concerned about the actual utilisation of these machines (are users querying all the time or just in bursts) and the cost of running such systems \cite{Omega13}. 

Comparing the actual scale or energy consumption of these systems is tricky and subject to many factors. Benchmarks are therefore needed to support a better standardisation and comparison of features for graph processing systems. In this regard, Graph500 and GreenGraph500 offer a starting point to measure and compare these aspects \cite{Graph500,GreenGraph500}.

\section{Conclusion}
\label{sec:conclusion}
We have reviewed the main graph analysis systems that can cope with billions of vertices/edges and revealed how there are just a few of them capable of processing near real-time queries when the topology of the graph changes. Streaming of the graph is the most commonly employed procedure for real-time processing. Dynamic repartitioning or replication and load balancing are the two main techniques used to adapt to changes in the topology of the graph. While repartitioning supports highly dynamic graphs and long-lived queries of a few users over a big portion of the graph, replication techniques serve best when there are short-lived queries of millions of users on small areas of the graph. More work is needed to understand the rates of graph changes and how these affect real world systems and algorithms. The survey led us to the BAR conjecture, which indicates that systems can have only two of the following features at the same time: big data, adaptability and real-time responses.

\section{Acknowledgments}

The authors would like to thank Alessandra Sala, Luis Rodero-Merino, Dionysios Logothetis and Claudio Martella for fruitful discussions and proofreading previous versions of this manuscript.

\end{document}